\documentclass[a4paper,10pt]{article}
\usepackage{mathrsfs}
\usepackage{graphicx}
\usepackage{latexsym}
\usepackage{amsmath}
\usepackage{amssymb}
\usepackage{textcomp}
\usepackage{amsbsy}
\usepackage{graphics}
\usepackage{epstopdf}
\begin{document}
 \title{\large \bf Lightest Kaluza-Klein graviton mode in a backreacted Randall-Sundrum scenario}
\author{Ashmita Das\footnote{E-mail address: ashmita@iitg.ernet.in} \\
Department of Physics\\
Indian Institute  of Technology\\
North Guwahati, Guwahati - 781039\\
Assam, India\\
and\\
 Soumitra SenGupta\footnote{E-mail address: tpssg@iacs.res.in}\\
Department of Theoretical Physics,\\
Indian Association for the Cultivation of Science,\\
2A $\&$ 2B Raja S.C. Mullick Road,\\
Kolkata - 700 032, India.\\[10mm]}
\date{}
\maketitle
\begin{abstract}
In search of extra dimensions in the ongoing LHC experiments, signatures of  
Randall-Sundrum (RS) lightest KK graviton have been one of the main focus
in recent years. The recent 
data from the dilepton decay channel at the LHC has determined the experimental lower bound 
on the mass of the RS lightest Kaluza-Klein 
(KK) graviton for different choices of underlying parameters of the theory. 
In this work we explore the effects of the backreaction 
of the bulk scalar field, which is employed to stabilise the RS model, 
in modifying the couplings of the lightest KK graviton
with the standard model (SM) matter fields located on the visible brane.
In such a  modified 
background geometry we show that the coupling of the lightest KK graviton 
with the SM matter fields gets a significant suppression due to the 
inclusion of the backreaction of the bulk stabilising scalar field.
This implies that  the backreaction parameter weakens the signals from RS scenario in collider experiments
which in turn explains the non-visibility of KK graviton in colliders.
Thus we show that the modulus stabilisation plays a crucial
role in the search of warped extra dimensions in collider experiments. 
\end{abstract}
\newpage
\section*{Introduction}
Till date, the world of subatomic particles, 
are best described by the Standard Model (SM) of elementary particles.
The validity of the SM has been confirmed with
a great accuracy in several experiments upto TeV scale.
The recent discovery of Higgs boson in Large Hadron Collider(LHC)
indeed is a major success story in this endeavour.
Such a successful theory however continues to encounter 
a longstanding but unresolved question in the context of 
the stability of the mass of Higgs boson against large
 radiative correction, known as gauge hierarchy/fine tuning problem.
Two most popular models, proposed in the context of this problem, are 
Supersymmtery and extra-dimensional models \cite{ADD,Antoniadis,Horava,RS1,RS2,ADD1,Kaloper,Cohen}. 
In absence of any signature of supersymmetry near TeV scale so far, the  
significance of the presence of extra dimension continues to grow. 
Among these models the warped geometry model proposed by Randall and Sundrum\cite{RS1} 
assumed a special significance because  a) it resolves the gauge hierarchy problem 
without introducing any other intermediate scale in the theory,
b) the modulus of the extra dimension can be stabilised by introducing
 a bulk scalar field\cite{GW1}  without any unnatural fine tuning of the parameter of the model.\\
 It  may also be mentioned that a warped solution, though not exactly same as RS model, can be found
from string theory which as a fundamental 
theory predicts inevitable existence of extra dimensions \cite{Green}.\\

Due to these features, detectors in LHC are designed
to explore possible signatures of the warped
extra dimensions through various decay channels
of RS Kaluza-Klein (KK) graviton. While CMS detector searches the signal of extra dimension through the  
final states of the decay into   
leptons and hadrons,  ATLAS detector is designed to capture the dileptonic decay of the KK gravitons.\\

\section*{\small Brief description of RS model}
The RS model is characterized by the non-factorisable 
background metric,
\begin{equation}
ds^2 = e^{- 2 k|y|} \eta_{\mu\nu} dx^{\mu} dx^{\nu} -dy^2 \label{eq1}
\end{equation}
 The extra dimensional coordinate is denoted by  $y$ and  ranges from
$-r_0$ to $+r_0$ following a $S^1/Z_2$ orbifolding.
Here, $r_0$ is the compactification radius of the extra dimension. Two 3-branes are 
located at the orbifold fixed points $y=(0,r_0)$. The standard model 
fields are residing on the visible brane and only gravity can propagate 
in the bulk. The quantity $k=\sqrt{\frac{-\Lambda}{24M^3}}$, which is of the order of 4-dimensional Planck
scale $M_{Pl}$. Thus $k$ relates the 5D Planck scale $M$ to the 5D cosmological constant
$\Lambda$.\\
The visible and Planck brane tensions are,
$V_{hid}=-V_{vis}=24M^3k^2$.
All the dimensionful parameters described above are
related to the reduced 4-dimensional Planck scale ${M}_{Pl}$ as,
\begin{equation}
 M_{Pl}^2=\frac{M^3}{k}(1-e^{-2kr_0})\label{rplanckmass}
\end{equation}
For $kr_0 \approx 36$, the exponential factor present
in the background metric, which is often called warp factor, produces 
a large suppression so that a mass scale of the order of Planck scale is reduced to TeV scale on the visible brane. 
A scalar mass say mass of Higgs is given as, 
\begin{equation}
 m_H=m_{0}e^{-kr_0}\label{physmass}
\end{equation}
Here, $m_H$ is Higgs mass parameter 
on the visible brane and $m_0$ is the natural scale of the theory
above which new physics beyond SM is expected to appear \cite{RS1}. \\
In RS model, the expressions for the mass of
first graviton KK mode $m_1$ and the coupling $\lambda$ with the SM matter fields
on the TeV brane as \cite{Davoudiasl},
\begin{equation}
 m_1=x_1 k e^{-kr_0}\label{gravmass1}
\end{equation}
where $x_1$ can be obtained from $J_1(x_n)=0$ \cite{Davoudiasl}, and 
\begin{equation}
 \lambda=\frac{e^{kr_0}}{M_{Pl}}\label{gravcoupling1}
\end{equation}
It has been argued in \cite{Davoudiasl} that the value of $k/M_{Pl}$ 
should be $0.1$ or less for the validity of classical 5-D solution for the 
metric in RS model. Keeping this constraint in mind, the 
ATLAS group in LHC estimated the lower bound on the mass of the lightest
KK graviton for different values of $k/M_{Pl}$. The absence of
KK graviton in dileptonic decay channels put stringent lower bound 
on KK graviton masses \cite{Atlas1,Atlas2}. According
 to the most recent experimental data \cite{Atlas2} at
$8$ TeV centre of mass energy and $20$ ${\rm{fb}}^{-1}$ luminosity, the
95\% confidence level lower limit on the RS lightest KK graviton mass
is further restricted to 2.68 TeV for $k/M_{PL}=0.1$.\\
We write eq(\ref{gravmass1}) as,
\begin{equation}
 m_1=x_1\frac{k}{M_{Pl}}M_{Pl}e^{-kr_0}\label{physmass1}
\end{equation}
From eq.(\ref{physmass1}), the mass of the 
RS lightest KK graviton can be tuned accordingly 
by increasing the warping parameter $e^{-kr_0}$ from $10^{-16}$,
so that it goes above the recent experimental lower bound proposed
by ATLAS for a fixed parameter $k/M_{Pl}$ which is related to coupling parameter
in the original RS scenario.\\
However from eq.(\ref{physmass}), it can be seen that if we increase the warping
parameter in order to raise the theoretically calculated graviton mass well
above the experimental lower bound then one needs 
to set the fundamental Planck scale of the theory ($m_0$)
a few order lower than the 4-D Planck scale ($M_{Pl}$) to obtain Higgs mass of
the order of $125$ GeV.
Therefore the increment of warp factor with the rise of this experimental lower bound 
on the mass of the RS lightest KK graviton, implies the 
inclusion of an intermediate energy scale in 
 between the Planck and TeV scale.\\
It has been mentioned earlier that the extra dimensional modulus in RS model can be stabilised to a value
of the order of inverse Planck length by introducing a massive
scalar field in the bulk \cite{GW1}. In this stabilising mechanism, the effect of the backreaction of the
bulk scalar field on the background geometry is neglected.
Later such warped geometry model was generalised by incorporating the back reaction of the
stabilising scalar field on the background metric \cite{Cline,wolfe,kribs,ssg1,ssg2}.
We therefore re-examine  the mass of
the lightest KK graviton and its coupling to the SM matter fields 
in such a modified  warped geometry model endowed with a
 back-reacted metric due to the stabilising bulk scalar field. 
In this work we demonstrate that due to the backreaction 
of the bulk stabilising scalar field on the background geometry,
the effective coupling of the lightest KK graviton with the SM matter fields 
becomes weaker, which in turn can explain the invisibility of RS lightest KK graviton
even if its lower mass bound is as low as few hundred GeV which is much below $2.8$ 
TeV as predicted by ATLAS.\\
Thus in this scenario we can explain the invisibility of KK graviton without
modifying the value of the warping parameter and $m_0$ from
their respective values in the original RS model.

We organize our work as follows:\\
In section(\ref{Genrs}), we describe five dimensional warped geometry model which includes the effect of the  
backreaction of the bulk stabilising scalar field on the background geometry.
Section(\ref{gravkkmode}) deals with the KK mass modes of graviton in this 
modified RS background. In section(\ref{grvint}), we discuss the lightest KK graviton interaction with 
the SM matter fields localised on the visible brane.
Sections(\ref{pheno1}) addresses the phenomenological implications and 
estimates the lower bound on the lightest graviton mass in the background of this back-reacted warped geometry model.
Section(\ref{conclusion}) ends with some concluding remarks. 
\section{\small Backreaction of the stabilising scalar field on the background geometry}
We consider the five dimensional action as, \cite{kribs}
\begin{eqnarray}
S=&-M^3\int d^{5}x\sqrt{g}R^{(5)}+\int d^{5}x\sqrt{g}(\frac{1}{2}\nabla \phi \nabla \phi-V(\phi))\nonumber\\
&-\int d^4x\sqrt{g_4}\lambda_{P}(\phi)-\int d^4x\sqrt{g_4}\lambda_{T}(\phi)\label{actionbr}
\end{eqnarray}
where $R^{(5)}$ is the five dimensional Ricci scalar, $\phi$ is the bulk scalar field and $V(\phi)$ is
the bulk potential term for the scalar field $\phi$, $g_4$ is the 
induced metric on the brane and $\lambda_P$, $\lambda_T$ are the potential terms   
on the Planck and TeV branes respectively due to the  bulk scalar field. 
The scalar field $\phi$ in general is a function of both $x^{\mu}$ and $y$. Here, we
consider the background VEV of the field $\phi(x^{\mu},y)\equiv \phi(y)$.
The 5-dimensional metric ansatz is \cite{kribs},\\
\begin{equation}
ds^2=e^{-2A(y)}\eta_{\mu\nu}dx^{\mu}dx^{\nu}-dy^{2}\label{mwarpfactor1}
\end{equation}
which preserve 4-D Lorentz invariance. The function $e^{-A(y)}$ is the modified warp factor.\\
As shown in \cite{kribs}, the 5-D coupled equations for the metric and the scalar field 
are:
\begin{equation}
 4A'^2-A''=-\frac{2\kappa^2}{3}V(\phi)-\frac{\kappa^2}{3}
\sum_{i}\lambda_i(\phi)\delta(y-y_i)\label{fieldeq1}
\end{equation}
\begin{equation}
 A'^2=\frac{\kappa^2\phi^{'2}}{12}-\frac{\kappa^2}{6}V(\phi)\label{fieldeq2}
\end{equation}
\begin{equation}
 \phi^{''}=4A'\phi^{'}+\frac{\partial V(\phi)}{\partial \phi}+\sum_i
\frac{\partial \lambda_i(\phi)}{\partial \phi}\delta(y-y_i)\label{scfieldeq1}
\end{equation}
where $\kappa$ is the five dimensional Newton's constant which is related to 
five dimensional Planck mass $M$ by $\kappa^2=1/(2M^3)$.
Here prime and $\partial_\mu$ denote the derivatives with respect to $y$ and 
4-D space time coordinate i.e.. $x^\mu$ respectively.\\
Following the procedure as illustrated in \cite{wolfe,kribs}, integrating equations
(\ref{fieldeq1}),(\ref{scfieldeq1}) on a small interval [$(y_i-\epsilon)$, $(y_i+\epsilon)$],
one finds the jump conditions,
\begin{equation}
 A'|_i=\frac{\kappa^2}{3}\lambda_i(\phi)\label{bdcA1}
\end{equation}
\begin{equation}
 \phi^{'}|_i=\frac{\partial \lambda_i(\phi)}{\partial \phi}\label{bdc2}
\end{equation}
As stated in \cite{wolfe}, to find the solutions for the above equations of motion
we actually need to reduce eq.(\ref{fieldeq1}-\ref{scfieldeq1}) to three
decoupled first order differential equations such that  two of them are separable.
The authors of \cite{wolfe} considered a 
definite form of the potential as,
\begin{equation}
 V(\phi)=\frac{1}{8}\left( \frac{\partial W(\phi)}{\partial \phi}\right) ^2-\frac{\kappa^2}{6}
W(\phi)^2\label{potform1}
\end{equation}
for some $W(\phi)$. \\
It is evident that if we implement  the two boundary
conditions [equations (\ref{condition1}, \ref{condition2})],
it solves both first order differential
equations 
$\phi'=\frac{1}{2}\frac{\partial W}{\partial \phi}$, 
$A'=\frac{\kappa^2}{6}W(\phi)\label{diffeq1}$ along with 
the Einstein and scalar field equations of motions in eqs.(\ref{fieldeq1}-\ref{scfieldeq1}).
\begin{equation}
 \frac{1}{2}W(\phi)|_{y_i-\epsilon}^{y_i+\epsilon}=\lambda_i(\phi)\label{condition1}
\end{equation}
\begin{equation}
 \frac{1}{2}\frac{\partial W(\phi)}{\partial \phi}|_{y_i-\epsilon}^{y_i+\epsilon}
=\frac{\partial \lambda_i}{\partial \phi}(\phi)\label{condition2}
\end{equation}
At this stage  we need to make a choice for $W$ to
solve for the back reaction of the bulk scalar field
on the metric. It has been shown in \cite{kribs} that 
inclusion of the back reaction of the stabilising field
generates a TeV order mass term for radion which may have 
interesting phenomenological consequences.\\
Considering the form of $W(\phi)$, chosen by the author of \cite{wolfe} and \cite{kribs}, 
\begin{equation}
 W(\phi)=\frac{6k}{\kappa^2}-u\phi^2\label{superpotential1}
\end{equation}
the brane potential terms become,
\begin{equation}
\lambda(\phi)_{+}=W(\phi_{+}) +
W'(\phi_{+})(\phi-\phi_{+})+
\gamma_{+}(\phi-\phi_{+})^2\label{branepot1}
\end{equation}
\begin{equation}
 \lambda(\phi)_{-}=W(\phi_{-}) +
W'(\phi_{-})(\phi-\phi_{-})+
\gamma_{-}(\phi-\phi_{-})^2\label{branepot2}
\end{equation}
Here +/- are used to represent Planck/TeV brane.
Choosing a definite form of $W(\phi)$, 
the solution for the stabilising scalar field ($\phi$) and the modified warp factor
$A(y)$ can be obtained as \cite{wolfe,kribs},
\begin{equation}
 \phi(y)=\phi_{P} ~e^{-uy}\label{solution1}
\end{equation}
\begin{equation}
 A(y)=ky+\frac{\kappa^2 \phi_{P}^{2}}{12}e^{-2uy}\label{warpsolution2}
\end{equation}
Here $r_0$ is the distance between two 3-branes which  can be stabilised
by matching the VEV  $\phi_P$ and $\phi_T$  of the stabilising scalar field $\phi$ at 
$0$ (location of the Planck brane) and $r_0$ (location of the TeV brane).
This implies  $ur_0={\rm ln}\left( \phi_{P}/\phi_{T}\right) $.
Therefore,
\begin{equation}
 e^{-ur_0}=\frac{\phi_T}{\phi_P}\label{brpotential}
\end{equation}
From equation (\ref{warpsolution2}), we observe that
the warp factor has modified from that in the  five dimensional Randall-Sundrum
model due to the backreaction of 
the stabilising scalar field. As expected, in the limit
$\kappa^2 \phi_{P}^2$, $\kappa^2 \phi_{T}^2\ll 1$ we retrieve the
original 5-D RS model.\\
All the dimensionful parameters described in this model are 
related to the reduced 4-dimensional Planck scale $\overline{M}_{Pl}$ as,
\begin{eqnarray} \label{rplanckmass1}
 &&M_{Pl}^2=\frac{M^3}{k}\bigg[
\left\lbrace 1-\left(\frac{\phi_P}{\phi_T}\right)^{-\frac{2k}{u}}\right\rbrace\\ \nonumber
&&-\frac{l^2}{3}
 \left(1+\frac{u}{k}\right)^{-1}
 \left\lbrace 1-\left(\frac{\phi_{P}}{\phi_{T}}\right)^{-2(1+k/u)}\right\rbrace \bigg]
\end{eqnarray}
where, $l=\frac{\kappa \phi_P}{\sqrt{2}}$\\
It was shown in \cite{kribs} that the factor $e^{-ur_0}$ appears in the final
expression for the radion mass which may have significant influence on
radion phenomenology.\\
Question that arises now : does the effect of the back reaction 
significantly modifies the KK graviton phenomenology also?
Can one explain the rise in the value of experimental lower mass bound for the lightest graviton KK mode
from the effect of the back reaction of the stabilising field?
We try to address this question in the following sections.
\section{\small lightest KK mass mode of Graviton in a back-reacted warped geometry} \label{gravkkmode}

The effective 4-D theory contains massless as well as massive KK tower of gravitons and all these higher excited states 
are coupled to the standard model fields, located on the TeV brane.
Our objective is to determine the mass of the first excited state of the 
graviton and its coupling with the SM matter fields in a 
back-reacted RS geometry due to the stabilising bulk scalar field.
In this context we wish to explore a possible explanation
for the hitherto non-visibility of the RS lightest KK graviton 
in the collider experiments.
The KK mass modes of graviton and its coupling 
with the SM matter fields in the background of the original RS model,
has been evaluated by the authors of \cite{Davoudiasl}. Here 
we extend the work by incorporating the back reaction of the bulk
stabilising scalar field.
The tensor fluctuations $h_{\alpha \beta}$ of the 
flat metric about its Minkowski value can be expressed through a linear expansion,
$\tilde{G}_{\alpha\beta}=e^{-2A(y)}(\eta_{\alpha\beta}+\kappa^{*}h_{\alpha\beta})$, where 
$\kappa^{*}$ is related to the higher dimensional Newton's constant. 
In order to find the graviton KK mass modes
we expand the 5-dimensional graviton field in terms of the
Kaluza-Klein mode expansion 
\begin{equation}
h_{\alpha\beta}(x,y)=\sum^{\infty}_{n=0}h^{(n)}_{\alpha\beta}(x)
\frac{\chi^{n}(y)}{\sqrt r_{0}}\label{grmmex}
\end{equation}
Where $h^{(n)}_{\alpha\beta}(x)$ are the KK modes of the graviton 
on the visible 3-brane and $\chi^{n}(\phi)$ are the corresponding 
internal wave functions for the graviton.
Imposing the gauge condition,
$\partial^{\alpha}h_{\alpha\beta}=0$ and compactifying  
the extra dimension, we obtain the effective 4-D theory for graviton
as,
\begin{equation}
 S_4=\int d^4x[\eta^{\mu\nu} \partial_{\mu}h_{\alpha\beta}^{(n)}(x)\partial_{\nu}
h^{\alpha\beta(n)}(x)-m_{n}^2h_{\alpha\beta}^{(n)}(x)
h^{\alpha\beta(n)}(x)]\label{eqngrav1}
\end{equation}
provided,
\begin{equation}
 \partial_y[e^{-4A(y)}\partial_y \chi^{n}(y)]+m_{n}^2 
e^{-2A(y)}\chi^{n}(y)=0\label{eqngrav2}
\end{equation}
and the orthonormality conditions
\begin{equation}
 \frac{1}{r_0}\int_{-r_0}^{+r_0} e^{-2A(y)}\chi^{n_1}(y) \chi^{n_2}(y) dy
= \delta_{n_1}\delta_{n_2}\label{orthonorm1}
\end{equation}
are satisfied.\\

Using $l=\frac{\kappa\phi_P}{\sqrt{2}}$ the warp factor
can be expressed as,
\begin{equation}
 A(y)=ky+\frac{l^2}{6}e^{-2uy}\label{modwarpfactor}
\end{equation}
For $l<\sqrt{6}$, we use a 
 leading order approximation for the series expansion of $e^{\frac{l^2}{6}e^{-2uy}}$.\\
 For $n=0$ {\it i.e..} zeroth mode of graviton, the differential equation for $\chi^{0}$
 turns out to be,
 \begin{equation}
  \partial_y[e^{-4A(y)}\partial_y \chi^{0}(y)]=0\label{eqngrav0mode}
\end{equation}
Solving the above differential equation and applying the continuity condition for the
graviton wavefunction at the two orbifold fixed points 
we obtain
\begin{equation}
\chi^0=c_1~=~{\rm constant}\label{solu0mode1}
\end{equation}
 Normalising the resulting wavefunction from 
eq.(\ref{orthonorm1}),  we finally get,
\begin{equation}
\chi^0=\sqrt{kr_0}\bigg[(1-e^{-2kr_0})(1-\frac{l^2}{3})\bigg]^{-1/2}\label{solu0mode}
\end{equation}
In order to find the solution for the higher KK graviton modes 
we define a set of new variables $\chi^{n}(y)=e^{2A(y)}\tilde{\chi}^{n}$
and $z_n=\frac{m_n}{k}e^{A(y)}=\frac{m_n}{k}e^{ky}(1+\frac{l^2}{6}e^{-2uy})$.
At $y=r_0$ the exponential series contains the factor
$e^{-ur_0}=\frac{\phi_T}{\phi_P}<1$, for $u>0$.\\
In terms of these new set of variables we obtain the following differential
equation for the graviton higher mode wave function,
\begin{equation}
 z_{n}^{2}\frac{d^2\tilde{\chi}^{n}}{dz_{n}^{2}}+z_{n}\frac{d\tilde{\chi}^{n}}{dz_n}
+\left[ z_{n}^{2}-4\right] 
\tilde{\chi}^{n}=0\label{eqngrav3}
\end{equation}
Solving the above equation we finally arrive at the solution for $\chi^{n}$,
\begin{equation}
\chi^{n}(y)=\frac{e^{2A(y)}}{N_n}\bigg[J_2\left(\frac{m_n}{k}e^{A(y)}\right) 
+\alpha_{n}Y_{2}\left(\frac{m_n}{k}e^{A(y)}\right)\bigg]\label{solwave1}
\end{equation}
where $N_n$ is the normalization constant for the wave function $\chi^n$. $J_2$, $Y_2$
are the Bessel function and Neumann function of order 2 and $\alpha_n$ is an
arbitrary constant. The KK mass modes of the graviton (i.e.. $m_n$)
and $\alpha_n$ can be found from the continuity condition
of the wave function at the two orbifold fixed points {\it i.e..} at
$y=r_0$ and $y=0$. The continuity condition at $y=0$ implies
$\alpha_{n}\ll1$ as $m_n/k\ll 1$. This leads to,
\begin{equation}
 \chi^{n}(y)=\frac{e^{2A(y)}}{N_n}J_{2}\left(\frac{m_n}{k}e^{A(y)}\right)\label{eqngrav5}
\end{equation}
The continuity condition at $y=r_0$, provides \\
\begin{equation}
 J_{1}(x_n)=0\label{ccondr_01}
\end{equation}
Where,
\begin{equation}
 x_n=\frac{m_n}{k}e^{A(r_0)}\label{ccondr_02}
\end{equation}
All these finally result into the expression for KK mass modes of graviton as,
\begin{equation}
 m_n=x_nke^{-A(r_0)}\label{mm1}
\end{equation}
The normalization constant $N_n$ for graviton wave function (\ref{eqngrav5}),
can now be determined by using 
the orthonormality condition in equation (\ref{orthonorm1}), as\\
\begin{equation}
 N_n=\frac{1}{\sqrt{kr_0}}
e^{A(r_0)}J_{2}(x_n)\label{normc1}
\end{equation}
\section{\small Coupling of the lightest KK graviton with standard model
matter fields on the visible brane}\label{grvint}

Let us consider the interaction of the first excited Kaluza-Klein
mode of graviton with the standard model matter fields residing on our universe i.e.. on the visible brane,
located at $y=r_0$. The solution for tensor fluctuations that 
appear on our visible brane can be obtained by
substituting the solution for $\chi^{n}(y)$ for $n = 0$ and higher modes ( see equ.(31), (34) and (38)) in equation (\ref{grmmex}),
at $y=r_0$,
\begin{eqnarray}
&&h_{\alpha\beta}(x,y=r_0)=\sum^{\infty}_{n=0}h^{(n)}_{\alpha\beta}(x)
\frac{\chi^{n}(r_0)}{\sqrt r_{0}}\\ \nonumber
&=&\sqrt{k}\left\lbrace\left(1-\frac{l^2}{3}\right)(1-e^{-2kr_0})\right\rbrace^{-1/2}
h^{0}_{\alpha\beta}(x^{\mu})\\ \nonumber
&+&\sum^{\infty}_{n=1}\sqrt{k}e^{A(r_0)}
h_{\alpha\beta}^{n}(x^{\mu})\label{finalsolngr2}
\end{eqnarray}
The interaction Lagrangian in the effective 4-D theory can be written as,
\begin{equation}
 \mathscr{L}|_{int}=-\frac{1}{M^{3/2}}T^{\alpha\beta}(x)
h_{\alpha\beta}(x,y=r_0)\label{coupling1}
\end{equation}
where $T^{\alpha\beta}$ is the energy-momentum tensor of the SM matter fields
on the visible brane and we use the relation between the 5-D Planck mass ($M_5$)
and the 4-D Planck mass ($M_{Pl}$) as shown in eq.(\ref{rplanckmass1}).\\
This leads to,
\begin{eqnarray}
&&\mathscr{L}|_{int}=-\frac{1}{M_{Pl}}T^{\alpha\beta}
h_{\alpha\beta}^{0}(x^{\mu})\\ \nonumber
&-&\frac{e^{A(r_0)}}{M_{Pl}}\bigg[\left\lbrace1-\left(\frac{\phi_P}{\phi_T}\right)^{-\frac{2k}{u}}\right\rbrace 
-\frac{l^2}{3}
 \left(1+\frac{u}{k}\right)^{-1}\\ \nonumber
&&\left\lbrace1-\left(\frac{\phi_{P}}{\phi_{T}}\right)^{-2(1+k/u)}\right\rbrace\bigg]^{1/2}
\sum^{\infty}_{n=1}T^{\alpha\beta}h_{\alpha\beta}^{n}(x^{\mu})\label{coupling2}
\end{eqnarray}
If we concentrate on the first excited KK mass mode of graviton
and its interaction with SM matter fields on the TeV brane,
the mass term can be identified as,
\begin{equation}
 m_1=x_1ke^{-A(r_0)}\label{mm2}
\end{equation}
while the interaction term of the first excited KK mode of graviton
with the SM matter fields on the TeV brane is,
\begin{eqnarray}\label{coupSM}
&& \Lambda|_{int}\cong \frac{e^{A(r_0)}}{M_{Pl}}
\bigg[\left\lbrace 1-\left( \frac{\phi_P}{\phi_T}\right)^{-\frac{2k}{u}}\right\rbrace -\frac{l^2}{3}
\left( 1+\frac{u}{k}\right)^{-1}\\ \nonumber
&&\left\lbrace 1-\left( \frac{\phi_{P}}{\phi_{T}}\right)^{-2(1+k/u)}\right\rbrace\bigg]^{1/2}
\end{eqnarray}
\section{\small Phenomenological implications}\label{pheno1}
\begin{figure}[!h] 
\begin{center}
\centering
\includegraphics[width=3.2in,height=2.3in]{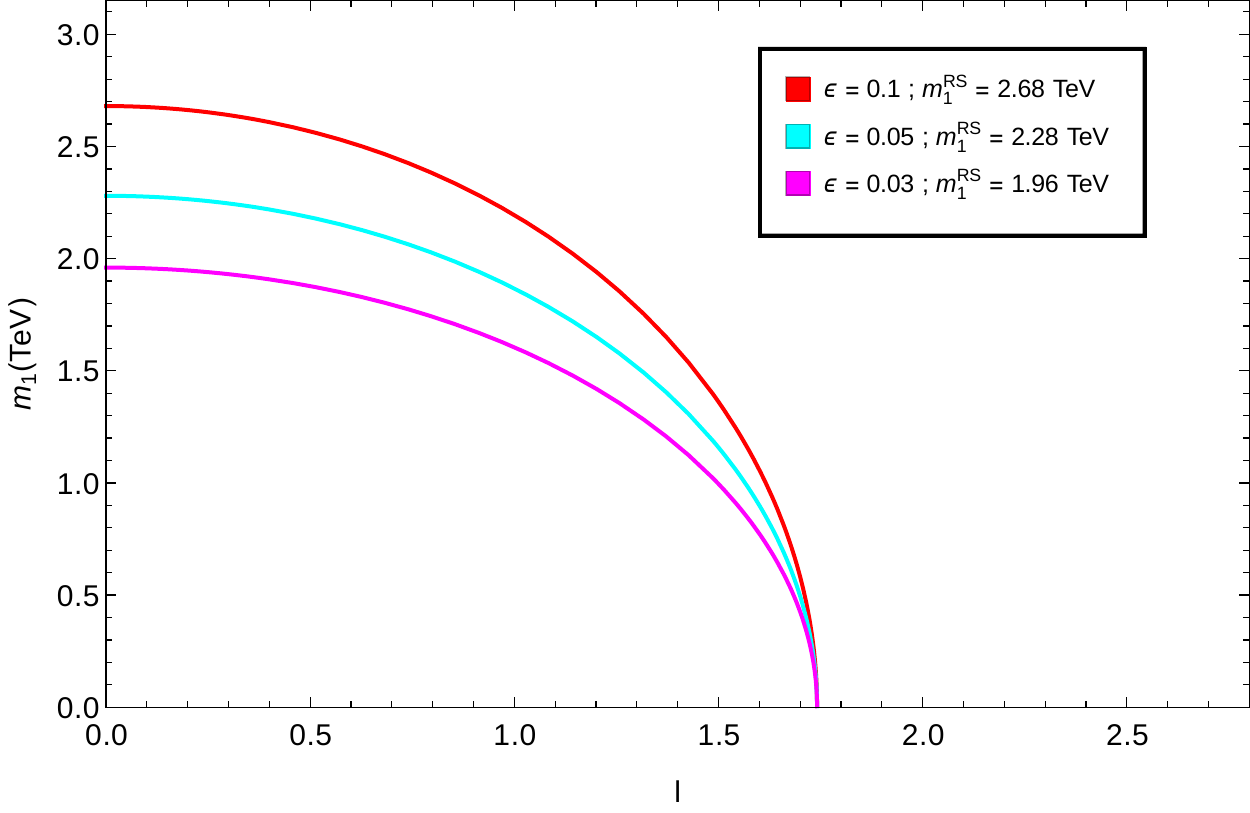}
\caption{Dependence of the new  mass bound of first  KK graviton on  parameter $l$ for
different choices of $k/M_{Pl}$}
\label{m1vsl}
\end{center}
\end{figure}

In the previous section we have given a description of the KK mass modes
of graviton and the interaction of the first excited KK mode of graviton with the SM
matter fields on the visible brane in the context of the back-reacted  RS model.
In eq.(\ref{coupSM}), we denote the term
\begin{eqnarray}\label{crossection3}
&&\bigg[\left\lbrace 1-\left( \frac{\phi_P}{\phi_T}\right) ^{-\frac{2k}{u}}\right\rbrace\\ \nonumber
&& -\frac{l^2}{3}\left( 1+\frac{u}{k}\right) ^{-1}
\left\lbrace 1-\left( \frac{\phi_{P}}{\phi_{T}}\right)^{-2(1+k/u)}\right\rbrace\bigg]^{1/2}
=\beta 
\end{eqnarray}
The parameter $\beta$ gives the modification of the coupling of KK graviton with the SM matter fields
from that evaluated in the original five dimensional RS model.\\
In order to address the gauge hierarchy problem we assume 
that the modified warp factor produces same warping as in the original RS scenario.
Therefore,
\begin{equation}
 A(0)-A(r_0)=-37\label{warping1}
\end{equation}
The above condition produces the following correlation among the parameters
$l$, $k/u$ and $\frac{\phi_P}{\phi_T}$:
\begin{equation}
 \frac{k}{u}=\frac{1}{{\rm ln}(\frac{\phi_P}{\phi_T})}
\bigg[37+\frac{l^2}{6}(1-\frac{\phi_{T}^{2}}{\phi_{P}^{2}})\bigg]
\label{warping2}
\end{equation}
The eq.(\ref{warping2}) dictates that for a particular choice of $\phi_P/\phi_T$ and $l$,
fixes the value of the parameter $k/u$  the value of $l<\sqrt{6}$ and $\phi_P/\phi_T>1$. We explore the parameter space by varying 
the backreaction parameter $l=1.68, 1.7, 1.71..$, and for each $l$
by varying $\phi_P/\phi_T=1.5, 2.5, 3.5..$
one can obtain the corresponding values of $k/u$ from eq.(\ref{warping2}).
After that we evaluate the parameter $\beta$ from eq.(\ref{crossection3}), which
varies over the values $0.17, 0.20, 0.22...$ corresponding to our 
different choices of the parameters of the model.\\
The values of $\beta$ clearly points out that there is a significant amount of
suppression in the dilepton decay channel of the lightest KK graviton over that 
evaluated in the original RS scenario. This implies that for appropriate 
choice of the parameters, the 
effect of back reaction of the bulk stabilising scalar field
on the background geometry of a warped extra dimensional model
can effectively suppress the coupling parameter of the lightest KK graviton with the SM matter fields.
This in turn reduces the value of the lower bound on the 
the mass of the lightest KK graviton.
For example , the  lower bound on the mass of the RS lightest KK
graviton, say for $\frac{k}{M_{Pl}}=0.1$
now  can be substantially lower than $\sim 2.8$ TeV ( lower mass bound without back reaction ) for appropriate choice of the parameter $l$.\\
Fig.\ref{m1vsl} clearly brings out  the dependence of the lower mass bound of first KK mode of graviton
with parameter $l$, for different choices of $k/M_{Pl}$ which indicates a significant suppression from the lower mass bound proposed by ATLAS for the original RS model.\\
We fix $\phi_P/\phi_T=1.5$ and write $\beta$ in terms of $l$  by replacing $k/u$
from  eq(\ref{warping2}). We then plot modified lower mass bound of first excited KK mode of graviton with 
$l$.
\newpage
 In summary, the  modulus stabilisation mechanism effectively reduces
the lower bound of the mass of the lightest KK graviton by a factor $\beta$
 which for appropriate choice of the parameter
values can be five-ten times lower than that in the original RS scenario.
\section{Conclusion}\label{conclusion}
We consider a generalised version of RS model where the effect of the backreaction 
due to the stabilising bulk scalar field on the background spacetime has been taken 
into consideration. We aim to study the contribution of this backreaction on the 
mass of the lightest KK graviton and its couplings to the SM matter fields.\\
Since the modulus stabilisation in braneworld model
 is an important requirement to make the prediction of the model more robust, 
 it is therefore worthwhile to look for experimental supports for the model in
 its stabilised version. 
Our study  strongly suggests
that due to the inclusion of the back reaction 
of the stabilising scalar field, the estimated value of the lower bound of the 
mass of the lightest KK graviton by the ongoing collider experiments
($m_1=\sim 2.8$ TeV for $k/M_{Pl}=0.1$),
may get reduced by approximately five-ten  times for a fixed $k/M_{Pl}$.
In summary, the backreaction of the bulk stabilising scalar field 
inevitably suppresses the
lower bound of the mass of the lightest KK graviton 
implying that there is no requirement to fine
 tune any parameter like warp factor or $m_0$ (natural scale of the theory)
to justify the estimated lower bound on the mass of RS lightest KK graviton
from the ongoing collider experiments.

\subsection*{Acknowledgement}
We thank Sourov Roy, Shankha Banerjee and Srimoy Bhattacharya for many illuminating discussions.

\end{document}